\documentclass[aps,prl,reprint,superscriptaddress,showpacs]{revtex4-1}
\usepackage{amsmath,amssymb,graphicx}
\newcommand{\unit}[1]{\ensuremath{\, \mathrm{#1}}}
\begin{document}

\title{Nodal Areas and Structured Darkness}

\author{Garreth J. Ruane}
\affiliation{Chester F. Carlson Center for Imaging Science, Rochester Institute of Technology, 54 Lomb Memorial Drive, Rochester, NY 14623, USA}
\author{Sergei Slussarenko}
\affiliation{Dipartimento di Fisica, Università di Napoli Federico II, Complesso Universitario di Monte S. Angelo, Napoli, Italy}
\author{Lorenzo Marrucci}
\affiliation{Dipartimento di Fisica, Università di Napoli Federico II, Complesso Universitario di Monte S. Angelo, Napoli, Italy}
\author{Grover A. Swartzlander, Jr.}\email{grover.swartzlander@gmail.com}
\affiliation{Chester F. Carlson Center for Imaging Science, Rochester Institute of Technology, 54 Lomb Memorial Drive, Rochester, NY 14623, USA}

\date{\today}

\begin{abstract}Generalized beams of light that are made to turn inside out, creating black nodal areas of total destructive interference are described.  As an example we experimentally created an elliptical node from a uniformly illuminated elliptical aperture by use of a lossless phase mask called a $q$-plate.  We demonstrate how a modified phase retrieval algorithm may be used to design phase masks that achieve this transformation for an arbitrary aperture shape when analytical methods are not available.  This generic wave phenomenon may find uses in both optical and non-optical systems.
\end{abstract}

\maketitle 
As a corollary to the principle that nature abhors a vacuum, it may be said that wave function amplitudes are disinclined to be zero-valued.  Exceptions are infinitesimally narrow nodal lines and points in the cross-section of a coherent wave.  The former occurs when the phase of the wave differs by $\pi$ across two regions, such as the zeros between Airy rings \cite{Airy1834} or when a wave has odd symmetry (e.g. $E(x>0,y)=-E(x<0,y)$).  Nodal points appear in the cross section when the phase azimuthally varies from 0 to $2\pi$ (or an integer multiple of $2\pi$) about the zero-valued point, forming a vortex wave \cite{Nye1974}.  An additional exception was discovered in the form of a circular nodal area \cite{Rouan2000,Mawet2005,Foo2005}.  The phase is singular in all these cases because it is undefined.  Zero-valued points in a system and their asymptotic limit are intriguing states of nature \cite{Berry2002} and attract considerable attention.  For example, laser speckle patterns are dense with vortices \cite{Baranova1981,Shvartsman1994}.  Wave function vortices may exhibit fluid-like dynamics in linear and nonlinear systems \cite{Landau1941,Ginzburg1958,Pitaevskii1961,Rozas1997}.  The dark structure of optical vortices enables advanced imaging techniques \cite{Furhapter2005,Westphal2005}, low roughness laser machining \cite{Hamazaki2010}, nano-scale photolithography \cite{Levenson2004}, and trapping of particles \cite{Ashkin1992,He1995,Gahagan1996,Arlt2000,Curtis2003}.  What is more, circular and spherical harmonic vortex wave functions are associated with spin and orbital angular momentum \cite{CohenTannoudji,Allen2003}.  In optics this enables encoding of information on a single photon \cite{Mair2001}, enhanced fiber or free space communication capacity \cite{Gibson2004,Wang2012,Tamburini2012,Bozinovic2013}, and an assortment of manifestations in quantum optics \cite{MolinaTerriza2007}.  Here we explore whether other nodal area shapes generally exist, beyond the circular case.  In particular we describe how light transmitted through a uniformly illuminated aperture may, in essence, be turned inside out by use of a lossless phase element.  Optical radiation protection and structured light illumination for various optics and photonics applications may be envisioned.  What is more, the ability to project a nodal area to a distant plane may find applications beyond the field of optics, such as electron waves or sound.

The purpose of this Letter is to introduce the basic concept of nodal areas and to pose currently unanswered questions about them.  We conjectured that any apertured beam of finite support may be optically transformed with a lossless phase element such that support of the final beam is the complement of the original.  To our knowledge a general analytical proof of this conjecture does not yet exist.  In fact, an analytical approach to find the matched functions of the phase element and the final beam containing a nodal area is an underdetermined problem.  However, we provide numerical examples showing single and disconnected apertures that have been transformed into nodal areas.  An analytical transformation that turns an apertured beam inside out was first realized for a circular aperture and was the basis for the construction of a high contrast coronagraph \cite{Rouan2000,Mawet2005,Foo2005}.  Wondering whether this was a unique manifestation, we investigated turning other apertured beams inside out \cite{Ruane2013}.  Here we explore how to form nodal areas using a lossless phase mask such as a so-called $q$-plate \cite{Marrucci2006} and experimentally verify that the circle is not the only case.  The discovery of other closed structures admitting nodal areas would be useful in optics for high contrast imaging applications, optical patterning, or light-matter interactions.  We point out that the zero-amplitude state is an idealized concept that in practice can only be approached asymptotically owing to limitations such as wave front aberrations, non-paraxial waves, the degree of coherence \cite{Swartzlander2007}, and quantum mechanics \cite{Berry2004}. 

\begin{figure}[t!]
\centerline{\includegraphics[width=\columnwidth]{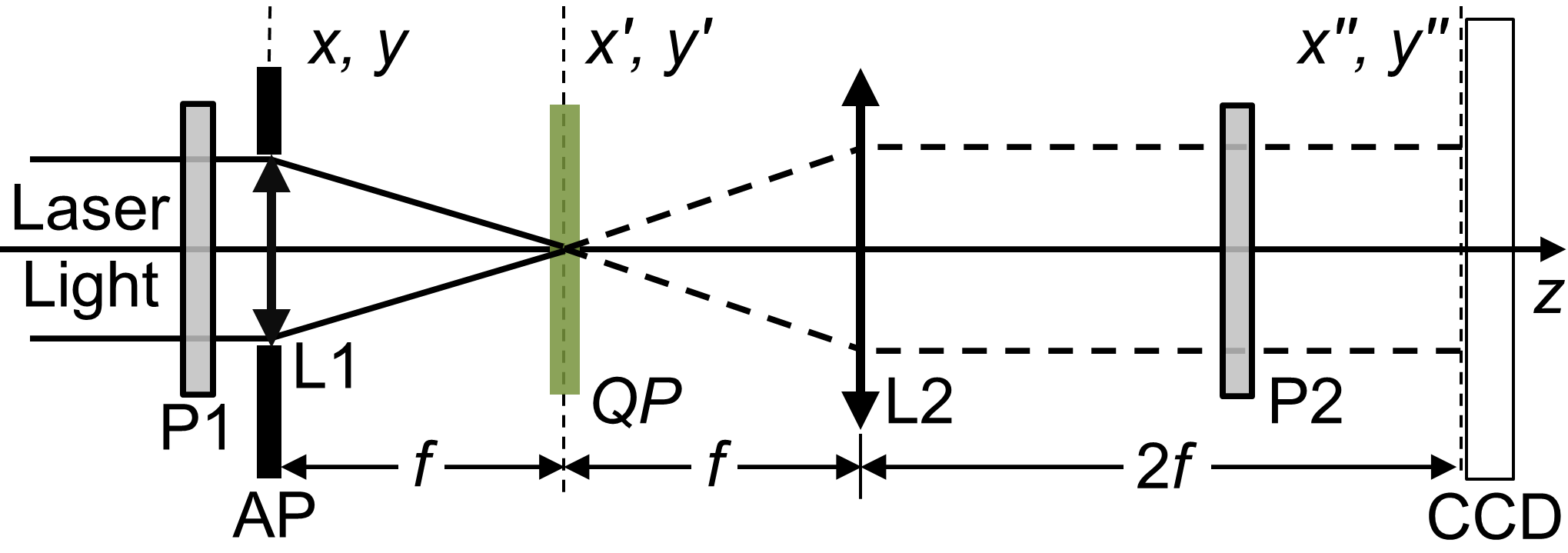}}
\caption{Optical configuration for producing a ring of fire with a $q$-plate (QP).   Uniform laser light that is circularly polarized by polarizing optics P1 enters an aperture AP at the $x,y$ plane.  Lens L1 focuses the transmitted light onto QP.  Lens L2 forms the exit pupil at the $x'',y''$ plane.  The second set of polarizing optics P2 is circularly cross-polarized to P1.  The ring of fire function appears at the $x'',y''$ plane.  A CCD detector array captures the image of the ring of fire.}
\label{fig1}
\end{figure}

A circular aperture of radius $R$ can be transformed into an isomorphic circular nodal area in a 4-$f$ optical system, illustrated in Fig. \ref{fig1} \cite{Mawet2005,Foo2005}.  The lens L1 focuses a uniform plane wave of incident light in the $x,y$ plane on a vortex phase element in the $x',y'$ plane.  Transmission of the optical field through the vortex element is represented by the function  where $m$ is a nonzero even integer called the topological vortex charge, and $\theta'$  is the azimuth in the transverse $x',y'$ plane (i.e. $\tan(\theta')=y'/x'$).  For experimental convenience we use a polarization-dependent element called a $q$-plate as the vortex element \cite{Marrucci2006}.  The transformed aperture function in the $x'',y''$ plane may be understood as a convolution of the circular aperture with the Fourier transform of the vortex transmission function \cite{Swartzlander2009}, $FT\{t_m(x',y')\}=a_m(1/r'')^2exp(im\theta'')$ where $r''$ and $\theta''$ are radial coordinates in the $x'',y''$ plane, $a_m=(-i)^{m+1}mf/k$, $f$ is the focal length of the lenses, $k=2\pi/\lambda$, and $\lambda$ is the wavelength.  Rather than imaging the aperture, this system remarkably diffracts all the light outside the aperture, resulting in a nodal area where $E(x'',y'')=0$ in the interior $(r''<R)$, and a ``ring of fire" in the exterior $(r''>R)$.  In the case where $m=2$, the exterior ring of fire function is $E(x'',y'')=(R/r'')^2\exp(i2\theta'')$.

A generalized circular nodal area is expected for any linear superposition of vortex transmission functions having nonzero even values of the topological charge \cite{Swartzlander2009}.  Such a superposition, however, does not change the circular size or shape of the ring of fire.  We questioned whether different aperture shapes could produce isomorphically similar rings of fire.  To our knowledge it is currently unknown whether the desired lossless transmission functions mathematically exist.  In fact, it is difficult to imagine how other apertures and phase transmission functions could also produce a nodal area based on either a convolution or direct integral point of view.  We may, however, formulate a phase retrieval approach to calculating the required transmission function.  Consider a 4-$f$ imaging system with aperture function $A(x,y)$ and transmission function $t(x',y') = \exp[i\Phi(x',y')]$.  The field at the focal plane just before the phase element is given by the Fourier transform of the aperture $F(x',y')=FT\{A(x,y)\}$.  Similarly, the exit pupil field is $G(x'',y'')=FT\{F(x',y')\exp[i\Phi(x',y')]\}$.  We wish to calculate the phase function $\Phi(x',y')$ necessary to form an exit pupil field $G(x'',y'')$ that is zero valued over the support of $FT\{F(x',y')\}$.  We were unsuccessful at finding a mathematical mapping that maintains phase-only transmission for seemingly simple apertures such as a hyper-ellipse defined by the boundary $(x/a)^p+(y/b)^p=1$, where $p$ is a positive real number.  

An important aperture function from a practical point of view is an annulus because many telescopes have a Cassegrain design containing a central obscuration.  Reflecting microscope objectives also share this design.  The discovery of a nodal area for this system could provide a path toward the direct observation of exoplanets.  Various attempts to achieve extreme high-contrast astronomical imaging have met with limited success on Cassegrains \cite{Ruane2013,Mawet2011,Carlotti2011,Mawet2013}.  Assuming a rotationally symmetric annular aperture, we may constrain the phase transmission function to have azimuthal modal symmetry as in the circular case, but allow radial variation.  That is, the phase function may be written $\Phi=m\theta'+\mu(r')$,  where $\mu(r')$ is the rotationally symmetric component.  The exit pupil field becomes an $m$th order Hankel transform $G(r'',\theta'')=e^{im\theta''}H_m\{F(r')e^{i\mu(r')}\}$ and has the form $G(r'',\theta'')=g_0(r'')e^{i\nu(r'')}e^{im\theta''}$, where $g_0(r'')$ and $\nu(r'')$ are the rotationally symmetric components of the amplitude and phase, respectively.  We were unsuccessful in finding an analytic expression for $\mu(r')$, $g_0(r'')$, and $\nu(r'')$ in general.  Reverting instead to numerical methods, we found that a modified Gerchberg-Saxton (G-S) phase retrieval algorithm was suitable for predicting phase-only masks that were matched to a given aperture to produce arbitrary nodal areas \cite{Gerchberg1972}.  This method takes advantage of analytically inspired phase elements that diffract a substantial amount light outside of the exit pupil.  We have found that a ``helical axicon" phase element of the form $\mu(r')=\pm\alpha r'$, where $\alpha>2kR/f$ is a suitable initial condition for the optimization algorithm \cite{Khonina1992}.  The approximate ring of fire is the Fourier transform of the product of the initial phase function and the point spread function (PSF) of the optical system.  The light that appears within the geometric exit pupil is numerically set to zero and the field is inverse Fourier transformed.  This yields a new focal plane amplitude and phase.  The amplitude is replaced with that of the PSF and a new approximate ring of fire is computed via the Fourier transform.  This process is repeated until the power within the geometric exit pupil decreases to a predefined stopping condition.  The algorithm returns the focal plane phase that transforms the PSF into a nodal area at the exit pupil.  However, there is no guarantee that the calculated phase mask is amenable to existing fabrication techniques.  By constraining the G-S algorithm to a single modal symmetry, it is possible to reduce the complexity of the phase function.  

\begin{figure}[t]
\centerline{\includegraphics[width=\columnwidth]{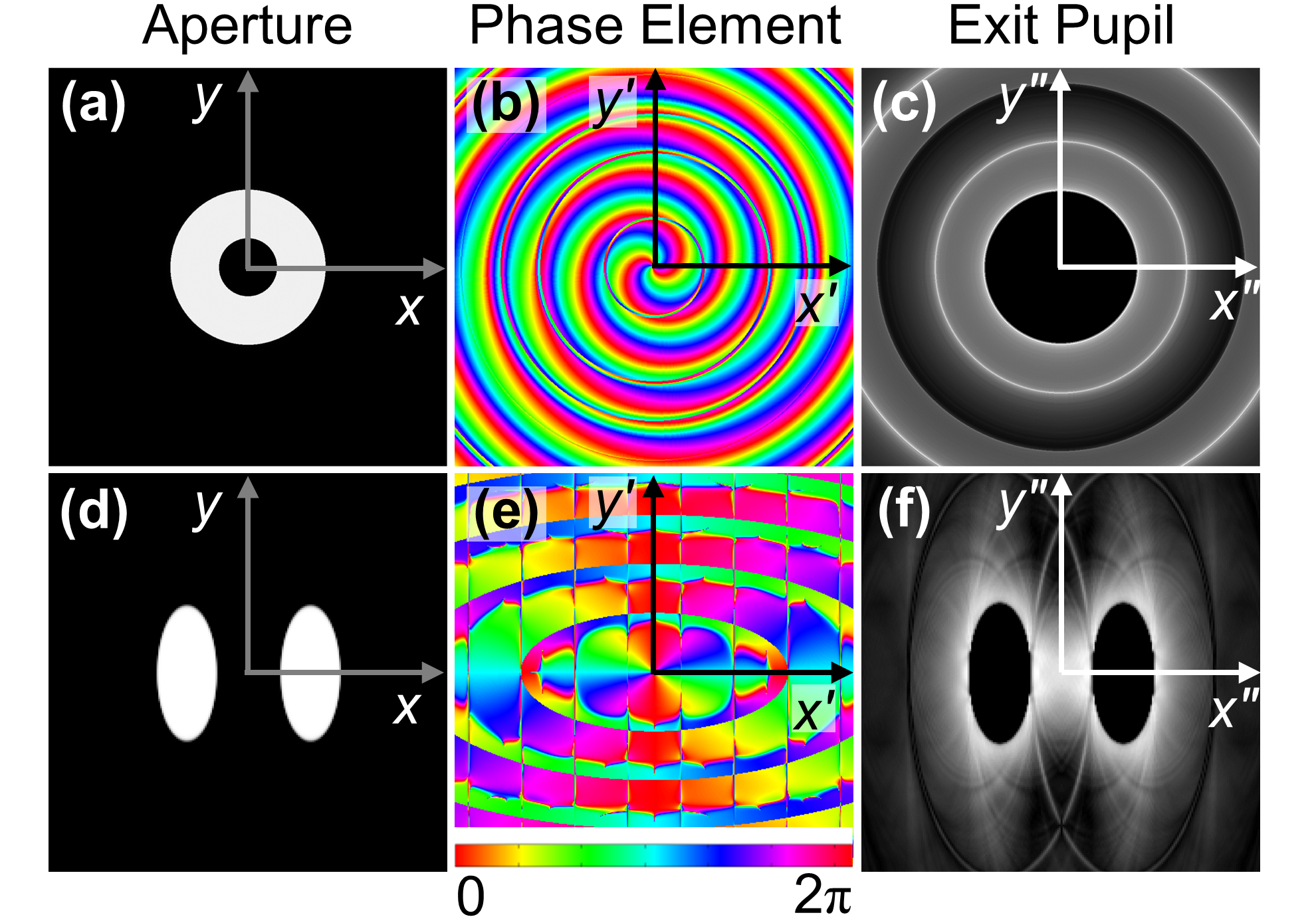}}
\caption{(a) An annular aperture matched with (b) the numerically computed phase mask generates (c) a ring of fire outside of the outer radius of the aperture.  (d)-(f)  Same as (a)-(c), but with a disconnected elliptical aperture pair, produced dual rings of fire (f).}
\label{fig2}
\end{figure}

An annular aperture, as well as the matched phase function of the vortex element and the predicted nodal area, are all shown in Fig. \ref{fig2}(a-c).  This, and other examples not reported here, tends to confirm our conjecture that a phase-only (lossless) mask can be found to produce a nodal area having the shape of an arbitrary entrance pupil.  We see in Fig. \ref{fig2}(b) that the phase function maintains the circular symmetry of the annulus, but that radial rings add complex structure (see Fig \ref{fig2}(a)).  The nodal area in Fig. \ref{fig2}(c) has the same diameter as the outer diameter of the annulus.  It too has complex radial structure not seen in the ring of fire for a circle.  These radial structures change for different values of the ratio of the outer and inner radii of the annulus (0.38 for Fig. \ref{fig2}(a)).

To further test our conjecture, we explored whether multiple disconnected apertures could also form nodal areas.  Multiple disconnected nodal areas could inspire a new branch of optics involving structured darkness.  As a first step in this direction we used the same procedure above, except the nodal area was defined as a pair of ellipses and the modal constraints were lifted.  Somewhat surprisingly, a phase-only mask that satisfied this condition could indeed be found, starting from an elliptical $m=2$ vortex as an initial guess in the G-S algorithm.  The aperture function, computed phase function, and nodal areas are shown in Fig. \ref{fig2}(d-f).  Again we find radial structure and other complicated patterns in the phase function.  The construction of such a phase element may be possible with lithographic patterning techniques \cite{Bomzon2002}.  Undoubtedly the computed structures are indicative of modal symmetries and relationships yet to be analytically discovered. 

\begin{figure}[b]
\centerline{\includegraphics[width=\columnwidth]{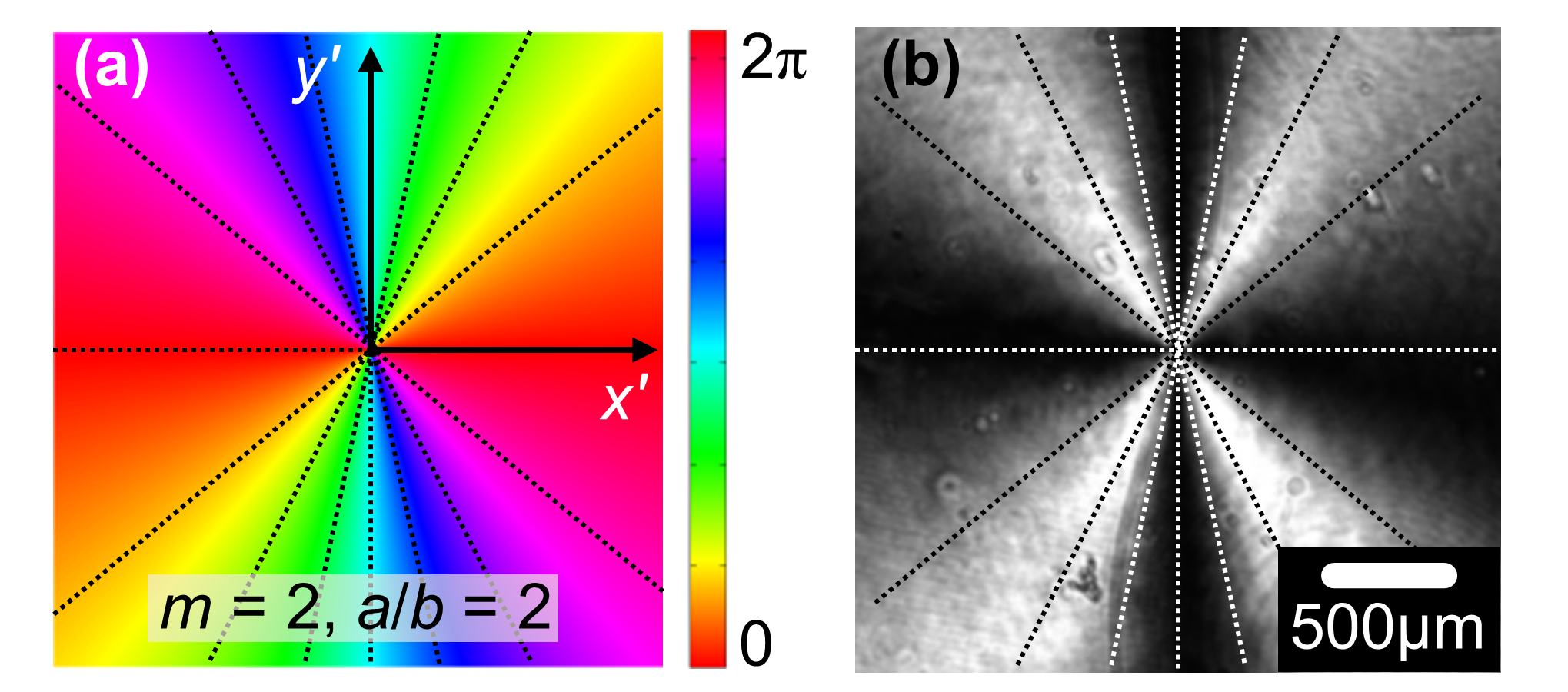}}
\caption{Transmission through an elliptical $q$-plate.  (a) Phase of the elliptical vortex function $\Phi(x',y')$ with $m=2$ and $a/b = 2$.  Lines of constant phase (dotted) are indicated in steps of $\pi/4$.  (b) Image of the fabricated elliptical $q$-plate placed between orthogonal linear polarizers, showing the expected irradiance pattern given by $\sin^2(2\alpha)$, where $\alpha$ is shown in steps of $\pi/4$.}
\label{fig3}
\end{figure}

The simplest aperture beyond a circle is an ellipse.  We found a skewed vortex transmission function that produces an elliptical nodal area and corresponding ring of fire: $\tilde{t}_m(x',y')=\exp(im\Phi)$, where $\Phi=\arctan(by'/ax')$, with the major and minor axes of the ellipse represented by $a$ and $b$ (see Fig. \ref{fig3}(a)) \cite{Ruane2013}.  We note that skewed vortices were explored in relation to an optical Magnus effect \cite{Rozas1999} and interesting propagation dynamics \cite{MolinaTerriza2001}.  A phase element producing the elliptical vortex function $\tilde{t}_m$ may be constructed by means of a $q$-plate, a spatially variant half-wave retarder with fast axis orientation described by $\alpha(x',y')=q\arctan(by'/ax')+\alpha_0$, where $q=|m/2|$ and $\alpha_0$ are constants \cite{Marrucci2006}.  The transmission of the $q$-plate may be written in terms of Jones matrices in the circular polarization basis as $E_{R,L}=e^{\pm i2\alpha}E_{L,R}$, where $E_R$ and $E_L$ are respectively the right and left hand circular polarization components of the incident field.  Note that each output vortex component is orthogonal to the input state \cite{Bomzon2002}.  The circularly polarized field components in the image plane of the aperture are found to be, for the case $m=2$
\begin{equation}
U_{R,L}\left(x'',y''\right)=E_{L,R}\frac{1}{{\rho ''}^2}\exp \left( \pm i2\Theta  \right),
\end{equation}
where $\rho''>1$, $\Theta=\arctan(ay''/bx'')$, and $\rho''=[(x''/a)^2+(y''/b)^2]^{1/2}$.  Both components are zero-valued if $\rho''<1$.   

The $q$-plate used in this demonstration was prepared by aligning a nematic liquid crystal (LC) material with the orientation angle $\alpha(x',y')=\arctan(by'/ax')$, i.e. $q=1$.  The desired planar alignment of the liquid crystals was induced using a photoalignment technique \cite{Chigrinov2008}.  We used 0.1\% wt. solution of Brilliant Yellow (BY) dye in dimethylformamide (DMF) as the aligning surfactant and Indium-Tin-Oxide (ITO) coated glass substrates so to have the possibility of applying an external electrical field to the LC film.  As part of the fabrication process \cite{Slussarenko2011,Slussarenko2013}, a polarized UV laser beam was expanded by a set of lenses, sent through a half-wave plate and focused on a sample with a cylindrical lens of 75\unit{mm} focal length.  Both the waveplate and the sample were attached to rotating mounts connected to a computer through step-motors.  By live control of the relative step size of two motorized mounts during exposure it was possible to impress elliptical orientation patterns with angular dependence.  Owing to the manufacturing process, a relatively small defect, 25\unit{\mu m} in diameter, occurs at the center of the sample, which is small compared to the dimensions of the $q$-plate ($20\unit{mm} \times 15\unit{mm}$).  

\begin{figure}[b]
\centerline{\includegraphics[width=\columnwidth]{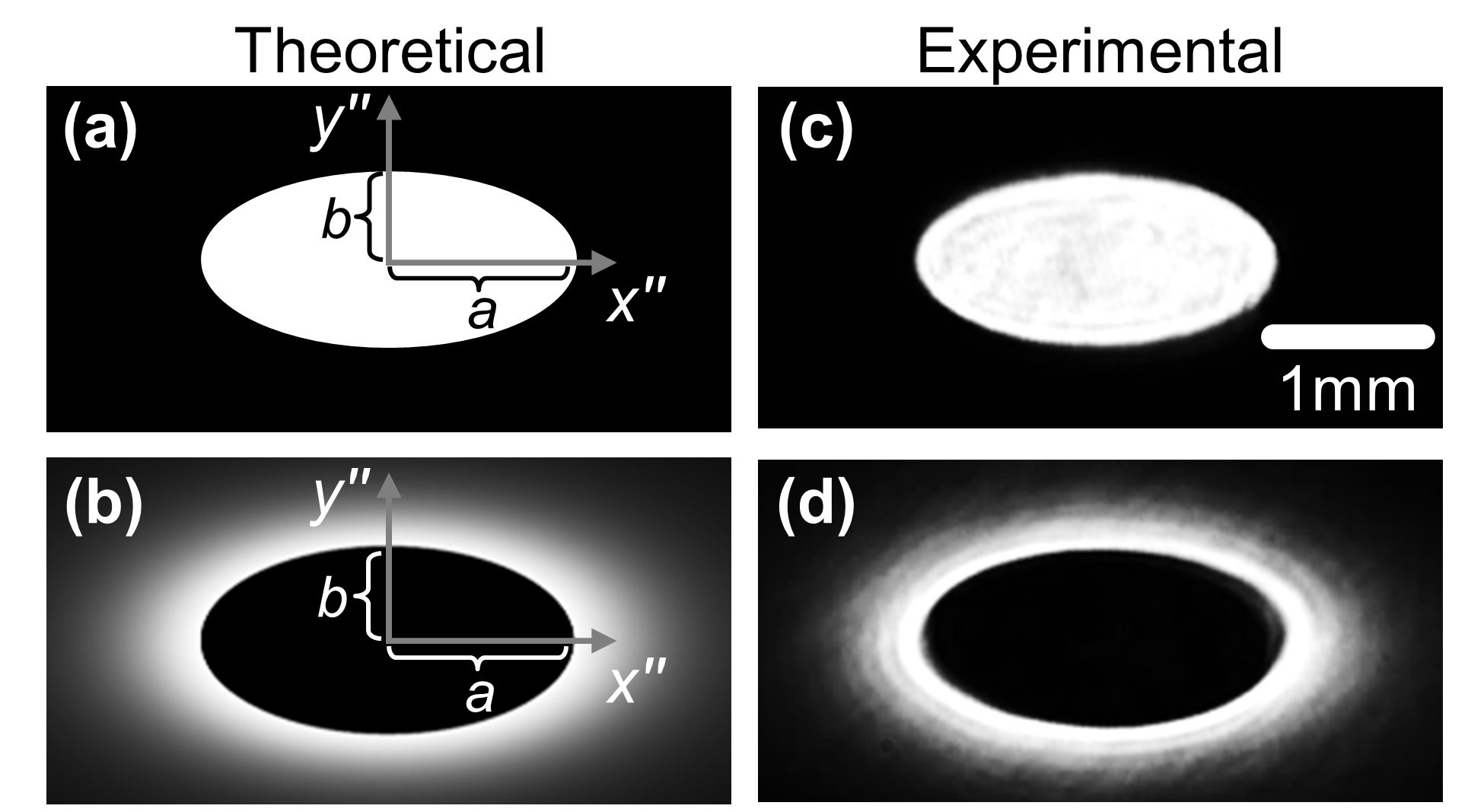}}
\caption{Experimental verification of an elliptical nodal area.  Theoretical beam profiles in the image plane of an elliptical aperture with (a) and without (b) the elliptical vortex $q$-plate.  Corresponding measured irradiance profiles (c) and (d), respectively.  The relative quality of the measure nodal area (d) is 0.0024.  The scale bar in (c) also applies to (d).}
\label{fig4}
\end{figure}

Figure \ref{fig3}(a) shows the expected phase pattern impressed on left-handed circularly polarized light; that is, an elliptical vortex function with topological charge $m=2$ and aspect ratio $a/b=2$.  When the $q$-plate is placed between two orthogonal linear polarizers, the expected transmitted irradiance, $\sin^2(2\alpha)$, agrees well with the experimental measurement shown in Fig. \ref{fig3}(b).  An elliptical ring of fire pattern was realized for the first time in the laboratory using the optical system illustrated in Fig. \ref{fig1} with the elliptical vortex $q$-plate.  An expanded, collimated, He-Ne laser beam ($\lambda=632.8\unit{nm}$) is made circularly polarized by use of a linear analyzer followed by a quarter wave retarder (P1).  An elliptically shaped beam having a uniform planar wave front was formed by transmitting the collimated beam through an elliptical aperture (AP) with major and minor axes $a = 1.50\unit{mm}$ and $b = 0.75\unit{mm}$ as depicted in Fig. \ref{fig4}(a).  This beam is expected to form a ring of fire with the light turned inside out, as in the theoretical image of Fig. \ref{fig4}(b).  The elliptical beam was focused by lens L1 onto the $q$-plate.  The LC was tuned to the half-wave phase retardation condition for wavelength $\lambda=632.8\unit{nm}$ by applying a $6.6\unit{V}$, $6\unit{kHz}$ voltage \cite{Slussarenko2011}.  The width of the focal spot was $515\unit{\mu m}$ along the $x'$ axis and $1030\unit{\mu m}$ along the $y'$ axis; that is, the focal spot is at least 20 times larger than the $\sim25\unit{\mu m}$ central deformation of the $q$-plate.  Lenses L1 and L2 both have a focal length of $1\unit{m}$, but different diameters: $25.4\unit{mm}$ and $50.8\unit{mm}$, respectively.  The field lens L2 imaged the entrance pupil on a CCD array in the $x'',y''$ plane.  An image of the entrance pupil forms on the detector array when the $q$-plate is removed from the system, or when the center of the $q$-plate is displaced from the center of the beam, as shown in Fig. \ref{fig4}(c).  However when these centers coincide, an elliptical ring of fire appears as seen in Fig. \ref{fig4}(d).  A circular polarizer analyzer (P2) was used to remove light that may have been transmitted in the wrong polarization state, owing to irregularities in the $q$-plate.  A dark frame was captured immediately before exposing the CCD detector array (SBIG STF-8300 cooled to $0.19^\circ\unit{C}$).  

A practical gauge of the quality of the experimentally measured nodal area is the fraction of the total beam power that is detected across the innermost half of the nodal region.  We achieved an experimental value of $2.4\times10^{-3}$, compared to the theoretical value of zero.

In conclusion, we have shown the experimental transformation of a uniformly illuminated elliptical aperture into a dark nodal area of equal size and shape using an elliptical vortex $q$-plate.  To our knowledge, this represents the first experimental demonstration of a noncircular nodal area.  In addition, our numerical results suggest that it is possible to form nodal areas of arbitrary shape with intricately designed phase elements.  The nodal areas are surrounded by a luminous ring of fire and may be made up of numerous detached shapes.  Structuring darkness in an optical field may be a valuable approach to light pattern control for various applications.  We anticipate that additional experimental and theoretical investigations will help uncover the mathematical properties of nodal areas as well as identify new applications of structured darkness.  

\begin{acknowledgments}
We thank Thomas Grimsley, and both Lindsay Quandt and Michael Rinkus, all from RIT for respectively fabricating the elliptical apertures and photographic work.  This work was supported by the U.S. Army Research Office under grant number W911NF1110333-60577PH and by the FET-Open Program within the 7th Framework Programme of the European Commission, under Grant No. 255914, Phorbitech.
\end{acknowledgments}

\end{document}